Preprint Notice:

# Exploring the contextual factors affecting multimodal emotion recognition in videos

Prasanta Bhattacharya, Raj Kumar Gupta, and Yinping Yang

**Abstract**— Emotional expressions form a key part of user behavior on today's digital platforms. While multimodal emotion recognition techniques are gaining research attention, there is a lack of deeper understanding on how visual and non-visual features can be used to better recognize emotions in certain contexts, but not others. This study analyzes the interplay between the effects of multimodal emotion features derived from facial expressions, tone and text in conjunction with two key contextual factors: i) gender of the speaker, and ii) duration of the emotional episode. Using a large public dataset of 2,176 manually annotated YouTube videos, we found that while multimodal features consistently outperformed bimodal and unimodal features, their performance varied significantly across different emotions, gender and duration contexts. Multimodal features performed particularly better for male speakers in recognizing most emotions. Furthermore, multimodal features performed particularly better for shorter than for longer videos in recognizing neutral and happiness, but not sadness and anger. These findings offer new insights towards the development of more context-aware emotion recognition and empathetic systems.

✦

## 1 INTRODUCTION

For long, the interest surrounding emotion recognition has been a key focus area within the field of affective computing [1]–[3]. Recent advances in computing infrastructure and datasets have led to new developments in detecting emotions using multiple types of data modalities like facial and acoustic expressions, linguistic and semantic patterns, body movements, eye gaze patterns and electroencephalography signals [2]. Recent studies have proposed different design approaches to perform automatic recognition of affective outcomes like valence, arousal, dominance, and emotion types [1], [2].

The ability to automatically recognize emotions has valuable implications for a range of applications. These include the design of empathetic agents [4], understanding consumer behavior at scale from observational data [5], and aiding healthcare professionals in a range of activities from diagnosing depressive symptoms in patients, to tracking the onset and progression of autism [6], [7].

This paper focuses on extracting high-level emotion features from visual, audio and text data modalities by leveraging a number of recent emotion analysis technologies. We tested the efficacy of these features in classifying fine-grained emotions from a real-world dataset comprising 2,176 labelled video clips, and sought to explore the conditions under which multimodal features outperformed bimodal and unimodal features.

More importantly, we performed an in-depth analysis of the effects from two important contextual factors that impact the performance of multimodal emotion recognition systems: i) *gender* of the speaker, and ii) *duration* of the emotion expression episode. First, we investigated whether the speaker's gender was linked to the relative effectiveness of visual, audio and text features in recognizing different emotions. To date, despite extensive research aimed at uncovering gender differences in how emotion is experienced and perceived (e.g. McDuff et al. [8]), little is known about how these differences affect the predictive performance of affective computing systems. We found that while multimodal features tend to outperform unimodal features for both male and female speakers, the relative improvement of using multimodal over unimodal was higher for male speakers. A trimodal classifier (i.e. combining visual, audio, and text features) was found to be the best performing multimodal classifier for both genders. For the unimodal features alone, we found that visual features performed best for both male and female speakers. With respect to the emotion categories in the overall sample, we found that while visual features performed best for predicting happy, sad, and neutral emotions, the audio features performed best for anger. Through these insights, the current study aims to be among the first to emphasize the importance of developing gender-aware multimodal systems, as the empirical performance of these systems can be largely contingent on inherent gender differences among speakers.

In addition to gender, we examined if the duration of the emotion being expressed affected the way a particular emotion was perceived and recognized across various modalities. Psychological studies in the past have examined the relationship between emotions and emotion duration using subjects' recalled emotional experiences [9]–[11].

- *P. Bhattacharya is with the Institute of High Performance Computing, Agency for Science, Technology and Research (A\*STAR), #16-16 Connexis North, 1 Fusionopolis Way, Singapore 138632 E-mail: prasanta_bhattacharya@ihpc.a-star.edu.sg.*
- *R.K. Gupta is with the Institute of High Performance Computing, Agency for Science, Technology and Research, #16-16 Connexis North, 1 Fusionopolis Way, Singapore 138632 E-mail: gupta-rk@ihpc.a-star.edu.sg.*
- *Y. Yang is with the Institute of High Performance Computing, Agency for Science, Technology and Research (A\*STAR), #16-16 Connexis North, 1 Fusionopolis Way, Singapore 138632 E-mail: yangyp@ihpc.a-star.edu.sg.*

However, to the best of our knowledge, the relationship between episodic duration of emotions and classifier performance in videos, as well as the associated underpinnings, are still not well understood. Our analyses show that duration played an important role in affecting classifier performance. Specifically, while multimodal classifiers outperformed unimodal classifiers for videos of all durations, this relative improvement was higher in shorter episodic durations for only the negative emotions like sadness and anger. In contrast, the relative improvement from using multiple modalities was higher in longer episodic durations for neutral and happiness emotions.

## 2 CONCEPTS AND RELATED WORK

Expressing and perceiving emotions is central to human experience and is a key contributor to the sustenance of interpersonal communication [12]–[14]. Emotions also affect how we interpret our environment, develop opinions and form judgments about the surrounding individuals and situations, and even drive behaviors [15].

Within the overarching category of 'affect', it is generally accepted that 'emotion' is distinct from related states like mood and temperament. Emotions might involve a range of feelings which may be transient, like relief, or last for relatively long periods of time, as is the case with mood, or really long spans of time, as with temperament [16]–[18]. Some of the earliest definitions about emotions came independently from psychologist William James and physiologist Carl Lange who contended that emotions arise naturally as a result of environmental stimulus, and that our interpretation of the physiological sensation of these changes constitutes emotion [19]. Contrary to James' contention that emotions could be a response to environmental changes, W.B. Cannon treated emotions as being felt first but expressed outwardly in certain behaviors much later [20]. There is also debate over how emotions can be perceived or described [13], [14], [21].

### 2.1 Emotion Type, Valence, Arousal and Intensity

For the purpose of this paper, we adopt three pertinent concepts from the emotion literature to facilitate the subsequent emotion feature extraction and analysis.

First, we refer to the stream of work on *categorization* of emotions which views emotions as discrete and distinct types such as happiness (or joy), anger, sadness and fear [22]–[26]. This view, while "simplistic"[27], [28], helps to focus the present study in terms of differentiating the specific effects due to data modality, duration and gender for various distinct emotion types. Second, we leverage the *dimensional* view of emotion that has received growing popularity in recent years [29]–[33]. In this study, we specifically considered two dimensions: valence and arousal. *Valence* describes the degree of pleasantness or unpleasantness of a signal and is generally measured on a continuous scale ranging from positive (i.e. pleasant) to negative (i.e. unpleasant). *Arousal*, on the other hand, refers to the extent of physical activation, ranging from no arousal to very high arousal. Thus, various emotional states like happy, angry, sad and afraid can be mapped to a two-dimensional valence-arousal scale, known as the circumplex model of affect [34]. For instance, in the circumplex model [34], "happy" is associated with a positive valence and moderate to high arousal. In contrast, both "tensed" and "angry" have a negative valence with moderate to high arousal, while "sad" is associated with a negative valence and low arousal. Third, we also consider emotion *intensity* as an important concept. This aspect of emotion concerns the degree or "depth" of emotional experience, ranging from barely noticeable to most intense imaginable [26], [35], [36]. For instance, the anger emotions can range from low-intensity annoyance to high-intensity rage, and this difference represents a relatively new, less explored aspect of emotional information that can benefit the development of computational methods (e.g., [37]).

### 2.2 Multimodal vs Unimodal Recognition Systems

Early research into emotion recognition systems largely focused on single data modalities (e.g., only facial expressions or tone of voice) and on emotions extracted from enacted sequences by trained actors [38]. More recently, studies have attempted to combine signals from multiple modalities such as facial expressions and audio [39], [40], audio and written text [41], [42], physiological signals [43], and various combinations of these modalities [44]. Overall, studies in this field have started to emphasize the importance of multimodal signals of emotions in more naturalistic scenarios (See [2], [3] for detailed reviews).

However, despite notable efforts, the existing literature on emotion recognition is limited in three ways. First, while there is growing evidence that multimodal systems generally outperform unimodal systems in relevant tasks [2], [40], [45], recent results have shown more complex patterns. For example, a number of studies have found that multimodal systems can often exhibit negligible improvements or even reductions in performance [39], [46]. Second, a vast majority of existing analyses on multimodal systems were based on relatively small samples (i.e. fewer than 50 participants), and using mostly bimodal classifiers. Third, there is also scant work on understanding the role of various kinds of contextual factors that might directly or indirectly influence model performance (e.g. de Gelder et al. [47]). With the exception of a few studies, the extant literature does not offer an adequate understanding of why certain data modalities work better for specific emotions, but not others [48]. It is therefore important to address the fundamental question of which data modalities are best suited for certain emotions, and the role of contextual factors in affecting the performance of these multimodal systems.

### 2.3 Contextual Factors: Gender and Duration

How does the performance of multimodal emotion recognition systems change in different contexts? A few recent studies, such as [39], [47], [49], [50], have started to explore how contextual factors can potentially affect the performance of emotion related classifiers. For instance, de Gelder et al. investigated how the presence of information on the visual and auditory context (e.g. information about natural scene, vocal expressions etc.) can benefit the recognition of facial expressions [47]. Contextual information can also come from the same modality. For instance, Metallinou et al. showed that emotions within as well as across utterances in a dialog sequence can be leveraged to improve emotion recognition [39]. These studies have focused primarily on *external* contexts (e.g., speaker's environment, emotion eliciting factors).

A relatively new direction is the interplay of *internal* contexts manifested through individual differences, like gender, and emotions. For example, Brebner analyzed two samples comprising Australian and International participants for differences in the frequency and intensity of self-reported emotions by gender [51]. The results highlighted significant differences for affection and sadness emotions, where females scored higher on both frequency and intensity. Males, on the other hand, scored higher on pride in both frequency and intensity. This is not entirely surprising, and a number of studies have attempted to explain gender differences in how emotions are experienced and perceived in others [8], [52], [53]. For instance, it has been shown that females generally perform better than men at perceiving and exhibiting negative emotions like sadness and fear [54]. The reasons for this range from evolutionary (e.g., Babchuck et al.'s primary caretaker hypothesis or Hampson et al.'s fitness threat hypothesis [52], [55]) to biological [54] to social and normative factors [56]. However, it remains to be understood whether and how the performance of multimodal emotion features varies as a result of such gender differences.

Another contextual factor that has attracted recent attention, and motivates the present study, is the role of *temporal* contexts. Recent studies have shown that the temporal length of the specific emotion being recorded is likely to influence the effectiveness of the classifier [48], [49]. This is likely due to a combination of three major factors. First, the duration of an emotion expression episode is likely to vary as a function of the data modality [49]. For example, Lingenfesler et al. [48] showed that an asynchronous fusion algorithm that takes into account different onset and offset times for emotions in various modalities, outperformed synchronous fusion algorithms. This is particularly useful since certain modalities, like language, spoken or written, take longer to manifest emotions, while others like facial expressions, take shorter time. Second, the nature of the platform or device through which the emotions are being recorded often influences the intensity and duration of the emotion. For instance, users recording a vlog on YouTube are likely to express emotions over longer durations, as compared to users uploading a much shorter Vine video. Lastly, the specific types of emotion might also be associated with the duration of the episode. In this study, we investigate how the duration of the videos affects the performance of multimodal features across various emotions.

## 3 DATA

### 3.1 Data Source

We make use of the One-Minute-Gradual (OMG) Emotion Dataset [57] for our analyses. The dataset was released as part of the 2018 OMG Emotional Behavior Challenge[1] where the original task was to implement an emotion-recognition system to accurately predict arousal and valence scores using an annotated dataset. The original OMG train dataset[2] contains YouTube links of 2,444 public video clips. We were able to retrieve 2,176 clips as of May 2018 using the links provided, as some links became unavailable over time. Each video clip's duration ranged around the one-minute mark in general, but with some shorter and longer duration videos. Most of the videos were of actors and actresses who were practicing lines or monologues probably in preparation for auditions. Through these monologues, the actors and actresses exhibited a number of discrete emotions (e.g., happiness, sadness, surprise), of varying arousals. Fig. 1 shows screenshots of six sample videos from this dataset.

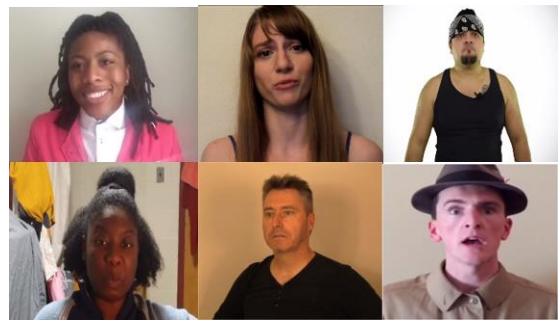

Fig.1. OMG-Emotion dataset (left to right, top to bottom: Actor/actress demonstrating happiness, sadness, anger, disgust, fear and surprise)

### 3.2 Emotion Ground Truth Labels

The dataset also provides utterance-level (approximately 10 seconds per utterance) majority-voted ground truth labels ("EmotionMaxVote") for seven discrete emotion categories, namely anger, fear, happiness, neutral, sadness, disgust, and surprise[3]. For each clip, the majority vote label was calculated by having five different annotations from crowdsourced Amazon Mechanical Turk workers [57].

Table 1 presents the basic descriptive statistics of emotion categories, sorted by the most balanced to the most imbalanced classes. We note that neutral and happiness had acceptable levels of imbalance, while sadness and anger exhibited higher class imbalance. Disgust, fear and surprise classes had the highest class imbalance. The relatively small number of disgust, fear and surprise cases in the dataset is an important limitation. It is for this reason that we choose to perform the subgroup analyses using only the neutral, happiness, sadness, and anger emotion categories.

---

[1] The OMG-Emotion Challenge: https://www2.informatik.uni-hamburg.de/wtm/omgchallenges/omg_emotion.html
[2] The data is downloaded from: https://github.com/knowledgetechnologyuhh/OMGEmotionChallenge/blob/master/omg_TrainVideos.csv

[3] The 2018 OMG Challenge required participants to predict valence and arousal scores only, and not discrete emotion categories.

TABLE 1
DESCRIPTIVE STATISTICS OF THE DATASET

| \#, count of videos %, out of the total of 2,176 videos | | | | | | |
|---|---|---|---|---|---|---|
| Neutral | Happiness | Sadness | Anger | Disgust | Fear | Surprise |
| 772 | 652 | 315 | 260 | 100 | 53 | 24 |
| 35.5% | 30.0% | 14.5% | 12.0% | 4.6% | 2.4% | 1.1% |

Here, it is useful to note that the class labels refer to the emotion ground truth of the speakers as *perceived* by the multiple independent annotators. It is different from *self-reported* or *recalled* emotional experiences, as used in traditional psychological studies.

### 3.3 Gender and Duration Labels

For this research, we manually annotated the gender of the speaker. We coded a discrete *gender* variable with three categories: "female", "male", and "others", for when it was either not possible to assess the gender or when there were multiple speakers in one utterance. We calculated the *duration* of each video using the start and end time of the video clips mentioned in the dataset.

## 4 EMOTION FEATURE EXTRACTION

As the purpose of our study is to understand the effects of contextual factors affecting multimodal emotion recognition, our feature extraction strategy focuses on extracting *theoretically explainable, high-level* emotion features from the visual, audio and text channels, respectively. Hence, even though it was possible to extract a large number of lower-level visual and non-visual features (e.g., facial action units, word embeddings), we chose to leverage a set of pre-trained emotion analysis systems that are grounded in data and key emotion concepts (see Section 2.1) to generate high-level emotion features.

### 4.1 Visual-based Emotion Feature Extraction

Among all modalities, facial expressions (e.g., frown, smiling) provide arguably the most intuitive emotion cues. We used a face emotion analysis system called *FEA*[4] to extract the visual features. FEA was trained with visual profiling techniques from static images [58], and was later extended to process a set of image frames in videos using deep learning techniques (see details in [58]–[62]). The system has been tested on AffectNet test set [63], and achieved Concordance Correlation Coefficients (CCC) of 0.527 and 0.569 for valence and arousal prediction respectively.

For each video, the FEA system generates *vValence* and *vArousal*, which indicate the degree of valence (from very unpleasant to very pleasant) and the degree of arousal (from no to very high physical activation) detected from the speaker's facial expressions. A third feature, *vIntensity*, measures the distance to the neutral state, calculated as the square root of the sum of squares of *vValence* and *vArousal*. The outputs also include 25 binary features (1: yes; 0: no) indicating whether, or not, a face is expressing a particular emotional state aligned with the circumplex emotion model [34]. Taken together, we extracted a total of 28 visual-based emotion features, namely *vValence*, *vArousal*, *vIntensity*, *vAfraid*, *vAlarmed*, *vAnnoyed*, *vAroused*, *vAstonished*, *vBored*, *vCalm*, *vContent*, *vDelighted*, *vDepressed*, *vDistressed*, *vDroopy*, *vExcited*, *vFrustrated*, *vGloomy*, *vHappy*, *vMiserable*, *vNeutral*, *vPleased*, *vSad*, *vSatisfied*, *vSerene*, *vSleepy*, *vTensed,* and *vTired*.

### 4.2 Audio-based Emotion Feature Extraction

In addition to visual features, the acoustic patterns of speech (e.g., speaking speed, pitch, intonation) provide different sources of emotion cues. For pre-processing, we first applied *FFMPEG*[5] to extract the audio component from the videos. We then extracted the audio-based emotion features using the *AcousEmo*[6] *system*, which was trained with acoustic analysis of emotion signals using deep Boltzmann machines (DBM) ([64]–[69]). The DBM classifier showed an effective predictive accuracy, measured as unweighted average recall, of 53.6, 64.4, 56.3 and 53.1 over valence, arousal, power and expectancy dimensions, respectively, when evaluated using the 2011 Audio/Visual Emotion Challenge dataset [70] (see details described in [69]).

For a given audio, AcousEmo produces two main outputs: *aValence* and *aArousal*, indicating the degree of valence (from very unpleasant to very pleasant) and the degree of arousal (from no to very high physical activation) detected from the acoustic signals. The system also generates two additional emotion dimensions: *aPower* (which subsumes two related concepts of the power and control that a speaker is expressing) and *aExpectancy* (which reflects whether the speaker is anticipating or expecting things to happen) (see theoretical discussion of the two additional dimensions [71] and the AVEC2011 challenge [70] for details). For the present OMG data, we extracted a total of ten audio-based emotion features*: aValence*, *aArousal*, *aPower*, *aExpectancy*, *aIntensity*, *aJoy*, *aAnger*, *aFear*, *aSadness*, *aNeutral*.

### 4.3 Text-based Emotion Feature Extraction

In addition to the visual and acoustic features, the language content of a speaker's speech in a video (e.g., word choice, linguistic meaning) can also carry valuable emotion cues. Before extracting the text-based emotion features, we first obtained the speech transcript from the audio using an Automatic Speech Recognition (ASR) engine developed by I2R[7]. This engine incorporates a range of speech processing techniques including sub-harmonic ratio based voice activity detection [72], mismatched crowdsourcing for acoustic modeling [73], multi-task learning for pronunciation modeling [74], and multi-task adversarial training for unsupervised adaptation [75], [76]. It has exhibited robust performance at various speech processing tasks including transcribing, retrieving, and indexing speech in English, Mandarin, Malay, Tamil, and other Southeast

---

[4] FEA is accessible via the authors [58]' page https://sites.google.com/site/vonikakis/software-code/appealing_slideshows or via https://opsis.sg (commercial version).

[5] FFMPEG is accessed via https://github.com/FFmpeg/FFmpeg
[6] AcousEmo is accessible by contacting the authors [65].
[7] The I2R ASR engine is accessible by contacting the authors [75].

Asian spoken languages [77]–[80]. The base engine we used is trained using a deep neural network (DNN), based on Kaldi speech recognition toolkit. Tested on dev-clean and test-clean conditions[8] of LibriSpeech data, the system achieved 9.32% and 9.49% word error rate (WER), respectively.

Next, we adopted *CrystalFeel*[9] to extract the emotion intensity features from the speech transcript. CrystalFeel is a collection of five SVM-based algorithms independently trained with tweets labelled with intensity scores for the overall valence, joy, anger, sadness and fear, respectively [37]. Evaluated on the SemEval-18 affect in tweets dataset [81], CrystalFeel achieved high Pearson correlation coefficients of 0.816 (overall valence), 0.708 (joy), 0.740 (anger), 0.700 (fear) and 0.720 (sadness) with human labelled emotion intensity scores [37]. CrystalFeel has also been useful in generating effective emotion features for predicting and understanding happiness in crowdsourced text [82] and predicting news popularity on Facebook [83].

For a given text such as a tweet or a speech transcript, CrystalFeel generates text-based emotion features in terms of the *intensity* value (from 0: barely noticeable to 1: extremely intense) of the five analytic dimensions. We were able to extract and use the following five text-based emotion intensity features from the speech transcript for each video: *tValence, tJoy, tAnger, tFear, tSadness*.

Details on the individual feature description and a bivariate correlational analysis, showing to what extent each feature is correlated with each of the OMG ground truth emotion labels, are provided in *Appendix A*.

## 5 EXPERIMENTS AND ANALYSES

In order to analyze and compare the effectiveness of different feature sets, we trained a support vector machine (SVM) with a polynomial kernel[10] and tuned the cost function using a grid search approach over the range of [0.01, 10]. This choice of classifier is consistent with recent work in this field (see [84] for an example).

For the classification models (i.e. classifying fine-grained emotion classes), the relevant accuracy measures considered were the raw accuracy and the area under curve (AUC). For the regression models (i.e. predicting arousal and valence scores), the relevant accuracy measure considered was the Pearson correlation coefficient.

### 5.1 Modality Effects: Comparing Unimodal, Bimodal and Trimodal Features

We first tested our multimodal features on the task of predicting continuous-valued arousal and valence scores provided by the OMG dataset. We developed a SVM regression model with a linear kernel and grid-search tuned cost function. Fig. 2 demonstrates the Pearson correlation coefficient from the regression models implemented for various feature combinations.

The results show that the trimodal model, denoted by *V+A+T*, performed the best for both valence and arousal.

Specifically, the best multimodal model outperformed the best unimodal model in Pearson correlation scores by over 23% and 20% for arousal and valence respectively. Interestingly, while the *V+A* model was found to be the most effective bimodal model in predicting arousal, the *V+T* and *A+T* models performed better in predicting valence. Similarly, while the audio-only (*A*) model was the best performing unimodal model in predicting arousal, the text-only (*T*) model performed best in predicting valence.

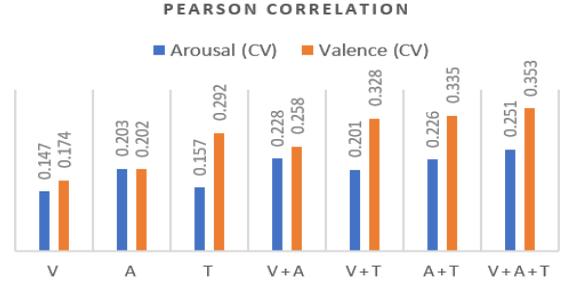

Fig.2. Pearson correlation coefficients for cross-validated (CV) models

We then evaluated the multimodal features in classifying the seven emotion categories using a tuned binary SVM, and a 10-fold cross-validation. We applied class weights proportional to the class distribution within each fold to account for the inherent class imbalance.

Table 2 below presents the raw accuracy and AUC scores from this analysis. Our results show that the bimodal and trimodal classifiers exhibited significant improvement in raw accuracy and AUC scores, over the unimodal classifiers to varying extents, across the different emotions. We tested the statistical significance of these improvements using a combination of a paired t-test as well as a Wilcoxon signed-rank test, and the differences were statistically significant at the 95% confidence level.

TABLE 2
RAW ACCURACY AND AUC RESULTS FOR EMOTION CLASSIFIER BASED ON 10-FOLD CROSS-VALIDATION

| Modality / Emotion | Neutral | Happiness | Sadness | Anger | Disgust | Fear | Surprise |
|---|---|---|---|---|---|---|---|
| | | | Raw Accuracy | | | | |
| Visual | **57.3** | **71.6** | 75.8 | 61.8 | 56.8 | 53.1 | 73.9 |
| Audio | 50.9 | 54.2 | 68.5 | 79.0 | 77.5 | **76.8** | **86.0** |
| Text | 42.1 | 66.6 | **84.3** | 79.5 | **85.3** | 43.1 | 44.4 |
| V + A | **62.0** | **74.9** | 76.2 | 76.5 | 82.6 | 72.9 | 89.6 |
| V + T | 59.5 | 71.1 | 77.8 | 65.6 | 72.9 | 63.0 | 78.6 |
| A + T | 56.6 | 59.1 | **78.2** | **79.7** | 89.2 | 86.4 | 92.5 |
| V + A + T | 64.5 | 75.6 | 82.4 | 80.9 | 88.3 | 84.1 | 95.6 |
| | | | AUC | | | | |
| Visual | .583 | .639 | .674 | .608 | .593 | .649 | .724 |
| Audio | .569 | .598 | .635 | **.651** | **.673** | **.725** | **.785** |
| Text | .536 | .598 | .599 | .571 | .581 | .598 | .616 |
| V + A | **.660** | **.695** | **.735** | **.720** | **.790** | **.842** | **.927** |
| V + T | .624 | .665 | .724 | .661 | .758 | .746 | .851 |
| A + T | .631 | .642 | .714 | .703 | .758 | .802 | .880 |
| V + A + T | **.688** | **.714** | **.799** | **.783** | **.872** | **.900** | **.937** |

---

[8] Kaldi training under different conditions: https://github.com/kaldi-asr/kaldi/blob/master/egs/librispeech/s5/RESULTS

[9] CrystalFeel is accessible via the authors' [37] page https://socialanalyticsplus.net/crystalfeel.

[10] The subgroup analyses were performed using a linear kernel.

To better quantify the improvements in accuracy, we use a version of the commonly used MM1 score which is specified as follows:

MM1=100*$[Max(a_{MM}) - Max(a_{UM})]/Max(a_{UM})$   (1)

Here, $Max(a_{MM})$ denotes the accuracy of the best multimodal classifier (*A+T* or *V+T* or *V+A* or *V+A+T*), $Max(a_{UM})$ denotes the accuracy of the best unimodal classifier (*A* or *V* or *T*). The effect size of the MM1 score has been used in a number of recent studies and offers a conservative estimate of the relative improvements from using a multimodal classifier [2]. Since MM1 measures improvement over a baseline score, the measure is not sensitive to the specific type of classification accuracy measure being used, and hence, offers a flexible and more disciplined way of comparing classifier performance across models.

Table 3 below presents the list of best unimodal and multimodal classifiers together with the associated MM1 effects. The estimates of MM1 scores show a number of interesting patterns. First, the average MM1 scores across the seven emotion classes is 20.2 for the cross-validated model, implying that the multimodal classifiers outperformed unimodal classifiers across all emotions. The highest improvements for multimodal classifiers were observed for the disgust and fear emotions.

TABLE 3
MM1 SCORES BASED ON 10-FOLD CROSS-VALIDATED AUC SCORES

| Modality | MM1 Score (%) | | | | | | |
|---|---|---|---|---|---|---|---|
| | Neutral | Happiness | Sadness | Anger | Disgust | Fear | Surprise |
| | 18.0 | 11.7 | 18.5 | 20.3 | 29.6 | 24.1 | 19.4 |
| *Best MM* | *V+A+T* | *V+A+T* | *V+A+T* | *V+A+T* | *V+A+T* | *V+A+T* | *V+A+T* |
| *Best UM* | *V* | *V* | *V* | *A* | *A* | *A* | *A* |

The best performing unimodal classifier for neutral, happiness and sadness was the visual-only (*V*) classifier, while the best performing unimodal classifier for the anger, disgust, fear, and surprise was the audio-only (*A*) classifier. Among the multimodal classifiers, the trimodal (*V+A+T*) classifier outperformed bimodal classifiers across all emotions. The best performing bimodal classifier across all emotion classes was the *V+A* classifier.

Given that disgust, fear, and surprise have high class imbalance, for rest of the analyses, we focus our attention on the emotion categories of neutral, happiness, sadness, and anger. We tested the robustness of the multimodal features using a random forest for these four emotion categories. The results from this analysis, detailed in Tables B1 and B2 in *Appendix B*, exhibit smaller effect sizes but are consistent with our key findings discussed here.

### 5.2 Contextual Effects: Gender and Duration

To enhance our understanding of the how gender of the speaker influences the inference of different emotions using a multimodal system, we performed an analysis of two gender categories (i.e. male vs. female). Using two human annotators, the dataset (2,176 videos) was gender-annotated into 738 male-only and 1,004 female-only videos. Around 434 videos could not be annotated with a specific gender as the gender of the speaker was either not discernible from the video even after inter-rater consultations, or was not uniquely identifiable on account of multiple speakers. For the successful gender annotations, the inter-rater agreement was 100%. Table B3 in *Appendix B* provides the sample distribution across gender and emotion categories.

Table 4 below highlights the MM1 scores for the male-only and female-only subgroups extracted from the data. Our results show that while multimodal classifiers offered significant[11] improvements over unimodal classifiers for both gender groups, there exists significant variance in the strength of this improvement, as discussed in detail in the next section. As mentioned earlier, we focus our attention only on the neutral, happiness, sadness, and anger emotions for this analysis, as the extent of class imbalance in these categories can be reasonably addressed using class weights and other techniques, e.g. SMOTE analysis [85]. Similar to the overall data, we tested the robustness of these findings for our gender subgroups using a random forest classifier. These additional results are presented in Tables B4 and B5 in *Appendix B*, and are consistent with our key findings discussed next.

TABLE 4
MM1 SCORES FOR MALE VS. FEMALE SPEAKERS BASED ON 10-FOLD CROSS-VALIDATED AUC SCORES

| Modality | MM1 Score (%) | | | |
|---|---|---|---|---|
| | Neutral | Happiness | Sadness | Anger |
| **Male Speakers** | 6.2 | 6.6 | 10.4 | 12.7 |
| Overall | 18.0 | 11.7 | 18.5 | 20.3 |
| *Best MM* | *V+A+T* | *V+A+T* | *V+A+T* | *V+A+T* |
| *Best UM* | *A* | *V* | *V* | *V* |
| **Female Speakers** | 1.7 | 4.8 | 5.8 | 11.1 |
| Overall | 18.0 | 11.7 | 18.5 | 20.3 |
| *Best MM* | *V+A+T* | *V+A+T* | *V+A+T* | *V+A+T* |
| *Best UM* | *V* | *V* | *V* | *V* |

---

[11] The difference in AUC scores between the best performing multimodal and unimodal classifier was statistically significant at 95% confidence level on a Wilcoxon signed-rank test.

The duration of an episode when an emotion is expressed can vary widely. As described earlier, the duration of episode might be longer or shorter depending on i) the modality e.g., emotions expressed through text take longer to express than through visuals, ii) the nature of medium e.g., short tweets and Vine videos vs. longer YouTube videos, and iii) the specific type of emotions, e.g., happiness takes shorter time to fully express than disgust.

In order to test the accuracy of the classifiers for the various emotion classes across short and long duration videos, we performed a quartile split on the training set based on the duration of videos, and categorized the first and fourth duration quartiles as short and long duration videos respectively. The first quartile comprised a total of 544 videos ranging in duration from 0.6 seconds to 4.4 seconds, while the fourth quartile comprised a total of 543 videos ranging in duration from 10.5 seconds to 30.7 seconds. Table B3 in the *Appendix B* illustrates the sample distribution across duration and emotion categories.

We then trained an SVM classifier, similar to the one in the previous section, on these two groups of videos. We report the resulting *MM1* scores for the two subgroups in Table 5 below. Similar to the gender subgroups, we tested and confirmed the robustness of this analysis using a random forest classifier; the additional results are presented in Tables B6 and B7 as part of *Appendix B*.

TABLE 5

MM1 SCORES FOR SHORTER VS. LONGER DURATION VIDEOS BASED ON 10-FOLD CROSS-VALIDATION

| Modality | MM1 Score (%) | | | |
|---|---|---|---|---|
| | Neutral | Happiness | Sadness | Anger |
| **Shorter Duration** | 0.0 | 6.3 | 10.3 | 17.9 |
| Overall | 18.0 | 11.7 | 18.5 | 20.3 |
| Best MM | V+A+T,V+A,V+T | V+A+T | V+A+T | V+A+T |
| Best UM | V | V | V | A |
| **Longer Duration** | 6.2 | 8.7 | 8.8 | 15.1 |
| *Overall* | 18.0 | 11.7 | 18.5 | 20.3 |
| Best MM | V+A+T | V+A+T | A+T | V+A+T |
| Best UM | V | V | V | V |

This analysis shows that multimodal classifiers outperformed unimodal classifiers across emotions for both shorter duration as well as longer duration videos. The only exception to this was the neutral class, where the best multimodal classifier trained on shorter duration videos performed at the same level as the best unimodal classifier (i.e. $MM1$ = 0). For neutral and happiness, classifiers trained on longer duration videos reported higher $MM1$ than those trained on shorter videos. However, for sadness and anger, classifiers trained on shorter duration videos reported a higher $MM1$ than those trained on longer videos.

## 6 DISCUSSIONS

### 6.1 Key Findings

**Multimodal emotion features perform significantly better than unimodal features, but the modality effects differ across emotion types**. One of the key insights from our analyses was the difficulty in classifying the neutral class. For the cross-validated model, the accuracy of classifying the neutral emotion was the lowest for the best unimodal (*V*, AUC=0.583), best bimodal (*V+A*, AUC=0.660) and trimodal (*V+A+T*, AUC=0.688) classifiers, as compared to all other emotions. Furthermore, the MM1 score ($MM1_{neutral}$ = 18.0) for the neutral emotion was lower than the average $MM1$ score across all emotions ($MM1_{avg}$ = 20.2). The same pattern is also consistent across all the gender (i.e. male vs. female) and duration (i.e. shorter vs. longer) subgroups, where the $MM1$ scores for neutral emotion was found to be smaller than the $MM1$ scores for other emotions. These findings imply that for classifying neutral states, multimodal classifiers do not offer substantial improvements over unimodal classifiers, as compared to other emotions. More importantly, this reduction in improvement was most salient for shorter duration videos where the best multimodal classifier showed no improvement over the best unimodal classifier ($MM1_{neutral}$ = 0). We found converging evidence for this using a random forest classifier as well ($MM1_{neutral}$ = 3.3).

This is consistent with recent work in affect recognition which emphasizes the complexity in conceptualizing what constitutes a neutral class and the problems with decoupling neutral-only videos from neutral segments of all other videos [84], [86]. Barros et al. reported that the neutral class exhibited the highest number of misclassifications in their analysis [86]. Similarly, Soleymani et al. reported high variance in pupillary response for participants when watching neutral scenes in videos [84]. Some past work has also shown that it is plausible for classifiers to confuse neutral state with emotions like happiness and sadness [87], [88]. In our analyses too, we found that the neutral and happiness emotion classes were the most difficult to classify, and showed lower improvements with multiple modalities as compared to other emotions. For all other emotions, multimodal accuracies were significantly higher than the unimodal accuracies for the training set.

Interestingly, we also observed that multimodal classifiers tend to work substantially better than unimodal classifiers for negatively valenced emotions like anger ($MM1_{anger}$ = 20.3 > $MM1_{avg}$ = 20.2), disgust ($MM1_{disgust}$ = 29.6 > $MM1_{avg}$ = 20.2), and fear ($MM1_{fear}$ = 24.1 > $MM1_{avg}$

= 20.2), although there is a possibility that these findings are specific to our data context, particularly given the imbalanced class distribution in disgust and fear categories. However, we found that, for both SVM and random forest classifiers, negative emotions like sadness and anger exhibited better MM1 scores than positively valenced emotions like happiness. This can be partly because these emotions are complex or compound [26] and are unlikely to be fully learned through any one modality. In contrast, emotions like happiness have clear markers in at least one modality (e.g., facial smile) and can therefore be learned with reasonable accuracy through unimodal classifiers. This is consistent with recent work by [89] which found that facial expressions conveyed happiness the clearest, while vocal features conveyed anger better than other emotions. In our analyses too, we found that the most effective unimodal classifier for the happiness emotion was visual-only (*V*, AUC = 0.639).

Similar to fine-grained emotions, the improvement in multimodal performance was substantial in predicting the continuous dimensions like arousal and valence. This is consistent with recent studies that have shown high efficacy of multimodal systems in predicting valence and polarity [1], [3], [84]. In our analyses, we found that across emotions, bimodal and trimodal regression models outperformed unimodal regression models for both arousal and valence. Among the multimodal models, the trimodal model (*V+A+T*) outperformed all bimodal models. Further, we found that while audio-related features (*V+A* and *A*) performed better than text-related features in predicting arousal, text-related features (*A+T, V+T, T*) performed better than audio features in predicting valence.

**The gender of the speaker plays a moderating role on the multimodal features' performance**. Our analyses on the OMG dataset also uncovered significant gender differences in emotion classification across various emotion classes. For instance, we found that multimodal classifiers outperformed unimodal classifiers for all emotions across the two genders (i.e. MM1 > 0). However, the average MM1 scores for male speakers ($MM1_{avg,male}$ = 9.0) outperformed those for female speakers ($MM1_{avg,female}$ = 5.9) across emotions. One possible explanation for this is that females, as young as 4 to 6 year old, have been found to express complex emotions [90], like sadness and anxiety, through a combination of subtle actions that are harder to detect, than for men who generally show limited reactions when expressing complex emotions. This is probably due to a mix of evolutionary limitations and restrictive social norms. For emotions like anger, where reactions like loud vocal expressions are relatively common for both men and women, the classifiers are less likely to show any difference. Recent studies looking at gender differences in emotion experience and perception have also found that women tend to be better than men at picking up certain emotions, and particularly negative emotions, as well as experiencing these emotions themselves [52], [53], [55]. However, our findings hint that, with the exception of anger, it might be challenging for multimodal classifiers to correctly classify other emotions expressed by female speakers.

Even though, in this study, we do not focus on gender effects in other negative emotions like fear and disgust due to the class imbalance in the available dataset, there have been recent studies that discuss relevant findings. For instance, and with disgust specifically, Al-Shawaf et al. provide a set of evolutionary-functional reasons to explain why women consistently experience higher levels of disgust than men [54]. It is possible, therefore, that female speakers might be expressing this emotion more vividly than males, making it easier for classifiers to learn this difference. Similarly, for anger, an emotion closely related to disgust, our analyses show that unimodal and multimodal classifiers perform reasonably well for female speakers. Furthermore, for male speakers, the visual-only classifier was found to be the best performing unimodal classifier for happiness, sadness, and anger emotions, while the audio-only classifier performed better for neutral. For female speakers, however, the visual-only classifier was found to be better performing for all emotion classes.

**Multimodal emotion features' effectiveness varies across episodic durations**. The other contextual factor that we focused on was the duration of the emotion expression episode. We conjectured that the duration of a recorded emotion episode would be contingent on i) data modality, ii) nature of the medium, and iii) the type of emotion. While certain modalities, like text, require longer to fully onset a specific emotion (i.e. we might need to wait for a sentence to end to decipher the full meaning and intent from the speaker), other modalities like visual are quicker and more direct (e.g. a smile to convey happiness). Videos from certain online video sharing websites like YouTube, as is the case with the current dataset, tend to be longer in duration than videos from other platforms like Vine.

Our analyses on classifier performance for shorter vs. longer duration videos show that the average *MM1* scores across emotions was comparable for the two duration groups ($MM1_{avg,shorter}$ = 8.6, $MM1_{avg,longer}$ = 9.7). However, we noticed variance across the specific emotion categories. For instance, and as mentioned earlier, multimodal classifiers failed to offer any improvement over unimodal classifiers for the neutral emotion when trained with *shorter* duration videos ($MM1_{neutral,shorter}$ = 0). For longer duration videos however, there was an improvement ($MM1_{neutral,longer}$ = 6.2). We also found that for neutral and positively valenced emotions, like happiness, the improvements in multimodal classification over unimodal classification were higher, as reflected by a higher *MM1* score[12]. However, for negatively valenced emotions like sadness and anger, the opposite was true. This might partly be due to the nature of the specific emotions being expressed. For instance, while negative emotions like sadness and disgust might take longer to fully onset and peak, others such as happiness and surprise tend to be more immediate. More research is needed to fully understand the underlying mechanisms of why these duration effects manifest.

---

[12] The improvements in AUC scores were statistically significant at the 95% confidence level on a Wilcoxon signed–rank test.

## 6.2 Implications, Limitations and Future Research

Taken together, our analyses provide clear and converging empirical evidence supporting the multimodal approach to recognizing and analyzing emotions. More importantly, our findings highlight significant variances based on important contextual factors, like the gender of the speaker and the duration of the emotion episode.

Notably, our findings add to a growing stream of literature that cautions about the prevalence of biases in machine learning datasets and models [91], [92]. Our findings highlight that the relative performance of emotion recognition systems might vary substantially across genders, durations, emotions, and plausibly, the underlying feature extraction systems. Researchers and developers need to be aware of such variances especially if such systems are being used in a gender-sensitive context (e.g. employee selection based on video interviews).

There are a number of limitations in the present study that warrants future research. First, the current study uses text, speech and visual as the three data modalities, and uses all of these to compare the performance of the multimodal emotion recognition systems. We note that the feature combination technique used is platform-agnostic, i.e., while these modalities are common sources of emotion data, the relative importance and richness of these modalities might vary across platforms, e.g. Instagram content is likely to be richer in visual than text.

Second, and as pointed out in [48], the various modalities are not temporally aligned within the duration of the emotion episode. This might introduce errors in feature fusion, leading to higher misclassifications.

Third, while our study offers robust analytical insights based on real-world videos, the findings can be limited to the specific dataset we used. As we noted, the OMG dataset, though with a relatively data size and a good gender and duration diversity, included relatively fewer representations of disgust, fear and surprise which prevented us from extending our subgroup analyses to these classes. We also did not have access to the annotators' demographic distribution, which prevented us from studying the sensitivity of our findings, if any, to the annotators' gender.

Last but not the least, in this study, we selectively focused on two important contextual factors: gender and duration. Other factors like i) surrounding scenes, ii) physiological state (e.g., body temperature), iii) conversational contexts, etc., may also affect classifier performance. Similarly, the results are also plausibly sensitive to the performance and biases associated with specific tools, though reasonably effective, that we selected to generate the features. The relative performance of the classifiers might, therefore, also vary as a function of the emotion features extraction technologies, and inherit their innate biases. As technologies advance, future work in this direction shall further explore the effects of these factors on multimodal classifier accuracy.

Future research can look into building richer datasets, applying more advanced emotion analysis technologies, and studying abstractions that can explain the psychological underpinnings driving the associations between emotions, data modalities, genders and other contextual factors. A theoretically sound interpretation of multimodal emotion recognition can benefit not just the development of better systems, but also help preempt biases and explain several inconsistent findings reported in prior work.

## 7 CONCLUSION

This study leverages state-of-the-art visual, audio and text based emotion recognition systems, and presents an empirical analysis to further our understanding of the contextual factors affecting multimodal emotion recognition. While multimodal features consistently outperformed unimodal features, our findings revealed significant variance in improvement across different emotions. This study is also an early attempt at examining the effect of contextual factors like the gender of the speaker or the duration of the emotional episode, on classifier accuracy across modalities and emotion classes.

From an applied perspective, the relative strength of the unimodal or bimodal features in comparison to full, multimodal emotion recognition can provide a valuable reference to practitioners. In today's digital platforms, emotion cues are largely extracted from text-based unimodal (e.g., Tweets, Facebook comments), audio-text bimodal (e.g., call center recordings), or visual-text bimodal (e.g., Instagram stories and posts) content. Practitioners can derive a sense of confidence from our multimodal analysis to make an informed assessment of the relative predictive performance of the unimodal and bimodal data sources.

Emotions will continue to play a key role in shaping human experience and communication. The exploratory findings from this study offer valuable empirical insights to help understand and inform how future affective systems leveraging multimodal emotions can be improved with multiple data and communication modalities, different emotions of interest, as well as a wide range of contextual factors including but not limited to gender and duration of the expressive episode.

## ACKNOWLEDGMENT

This research is supported by the Agency for Science, Technology and Research (A*STAR) under its SERC Strategic Fund (a1718g0046) and A*ccelerate Gap Fund (accl180222). We thank the Project Digital Emotions team for the valuable teamwork, and in particular, Nur Atiqah Othman for her annotation assistance on the gender labels used in the analysis. The authors are grateful for the helpful comments from Desmond Ong, and the advices from Stefan Winkler, Huang Dongyan and Nancy Chen on applying their systems in processing and extracting visual and speech features used in this study. All errors that remain are our sole responsibility.

## REFERENCES

[1] S. K. D'Mello, N. Dowell, and A. Graesser, "Unimodal and Multimodal Human Perception of Naturalistic Non-Basic Affective States during Human-Computer Interactions," *IEEE Trans. Affect. Comput.*, vol. 4, no. 4, pp. 452–465, 2013.

[2] S. K. D'mello and J. Kory, "A review and meta-analysis of multimodal affect detection systems," *ACM Comput. Surv.*, vol. 47, no. 3, pp. 1–36, 2015.

[3] R. A. Calvo and S. D'Mello, "Affect detection: An interdisciplinary review of models, methods, and their applications," *IEEE Trans. Affect. Comput.*, vol. 1, no. 1, pp. 18–37, 2010.

[4] B. Liu and S. S. Sundar, "Should Machines Express Sympathy and Empathy? Experiments with a Health Advice Chatbot," *Cyberpsychology, Behav. Soc. Netw.*, vol. 21, no. 10, pp. 625–636, 2018.

[5] H. Cui, V. Mittal, and M. Datar, "Comparative experiments on sentiment classification for online product reviews," in *Proc. of 21st National Conference on Artificial Intelligence, AAAI, MA*, 2006, pp. 1265–1270.

[6] M. Uljarevic and A. Hamilton, "Recognition of Emotions in Autism: A Formal Meta-analysis," *J. Autism Dev. Disord.*, vol. 43, no. 7, pp. 1517–1526, 2013.

[7] S. Alghowinem, R. Goecke, J. Epps, M. Wagner, and J. F. Cohn, "Cross-Cultural Depression Recognition from Vocal Biomarkers," in *INTERSPEECH*, 2016, pp. 1943–1947.

[8] D. McDuff, E. Kodra, R. el Kaliouby, and M. LaFrance, "A Large-scale Analysis of Sex Differences in Facial Expressions," *PLoS One*, vol. 12, no. 4, 2017.

[9] N. H. Frijda, B. Mesquita, J. Sonnemans, and S. Van Goozen, "The duration of affective phenomena or emotions, sentiments and passions," *Wiley Int. Rev. Stud. Emot.*, pp. 187–225, 1991.

[10] K. Brans and P. Verduyn, "Intensity and duration of negative emotions: Comparing the role of appraisals and regulation strategies," *PLoS One*, vol. 9, no. 3, p. e92410, 2014.

[11] P. Verduyn and S. Lavrijsen, "Which emotions last longest and why: The role of event importance and rumination," *Motiv. Emot.*, vol. 39, no. 1, pp. 119–127, 2015.

[12] C. Darwin and P. Prodger, *The Expression of the Emotions in Man and Animals*. USA: Oxford University Press, 1998.

[13] E. M. Zemach, "What is Emotion?," *Am. Philos. Q.*, vol. 38, no. 2, pp. 197–207, 2001.

[14] W. James, "What is an Emotion?," *Mind*, vol. 9, no. 34, pp. 188–205, 1884.

[15] G. L. Clore, K. Gasper, and E. Garvin, "Affect as Information," *Handb. Affect Soc. Cogn.*, pp. 121–144, 2001.

[16] S. P. Robbins and T. A. Judge, "Emotion and Moods," in *Organizational Behavior*, 17th ed., Pearson Education, Limited, 2017, pp. 138–171.

[17] J. M. Jenkins, K. Oatley, and N. Stein, *Human Emotions: A Reader*. Wiley-Blackwell, 1998.

[18] N. H. Frijda, "Moods, Emotion Episodes, and Emotions," *Handb. Emot.*, pp. 381–403, 1993.

[19] W. James, F. Burkhardt, F. Bowers, and I. K. Skrupskelis, *The Principles of Psychology*. London: Macmillan, 1890.

[20] W. B. Cannon, "The James-Lange Theory of Emotions: A Critical Examination and an Alternative Theory," *Am. J. Psychol.*, vol. 39, no. 1, pp. 106–124, 1927.

[21] J. Kagan, *What is Emotion?: History, Measures, and Meanings*. Yale University Press, 2007.

[22] M. Rashid, S. A. R. Abu-Bakar, and M. Mokji, "Human emotion recognition from videos using spatio-temporal and audio features," *Vis. Comput.*, vol. 29, no. 12, pp. 1269–1275, 2013.

[23] G. Krell *et al.*, "Fusion of Fragmentary Classifier Decisions for Affective State Recognition," in *IAPR Workshop on Multimodal Pattern Recognition of Social Signals in Human-Computer Interaction*, 2012, pp. 116–130.

[24] S. K. D'mello and A. Graesser, "Multimodal semi-automated affect detection from conversational cues, gross body language, and facial features," *User Model. User-adapt. Interact.*, vol. 20, no. 2, pp. 147–187, 2010.

[25] P. Ekman, "An Argument for Basic Emotions," *Cogn. Emot.*, vol. 6, no. 3–4, pp. 169–200, 1992.

[26] A. Ortony, G. L. Clore, and A. Collins, *The cognitive structure of emotions*. Cambridge university press, 1988.

[27] A. Ortony and T. J. Turner, "What's Basic About Basic Emotions?," *Psychol. Rev.*, vol. 97, pp. 315–331, 1990.

[28] T. J. Turner and A. Ortony, "Basic Emotions: Can Conflicting Criteria Converge?," *Psychol. Rev.*, vol. 99, pp. 566–571, 1992.

[29] S. Wang, Y. Zhu, G. Wu, and Q. Ji, "Hybrid Video Emotional Tagging using Users' EEG and Video Content," *Multimed. Tools Appl.*, vol. 72, no. 2, pp. 1257–1283, 2014.

[30] T. Baltrušaitis, N. Banda, and P. Robinson, "Dimensional affect recognition using continuous conditional random fields," in *2013 10th IEEE International Conference and Workshops on Automatic Face and Gesture Recognition (FG)*, 2013, pp. 1–8.

[31] M. S. Hussain, H. Monkaresi, and R. A. Calvo, "Combining Classifiers in Multimodal Affect Detection," in *Proceedings of the Tenth Australasian Data Mining Conference*, 2012, pp. 103–108.

[32] K. Lu and Y. Jia, "Audio-visual Emotion Recognition with Boosted Coupled HMM," in *Proceedings of the 21st International Conference on Pattern Recognition (ICPR2012)*, 2012, pp. 1148–1151.

[33] D. Glowinski, N. Dael, A. Camurri, G. Volpe, M. Mortillaro, and K. Scherer, "Toward a minimal representation of affective gestures," *IEEE Trans. Affect. Comput.*, vol. 2, no. 2, pp. 106–118, 2011.

[34] J. A. Russell, "A circumplex model of affect," *J. Pers. Soc. Psychol.*, vol. 39, no. 6, pp. 1161–1178, Dec. 1980.

[35] J. Sonnemans and N. H. Frijda, "The structure of subjective emotional intensity," *Cogn. Emot.*, vol. 8, no. 4, pp. 329–350, 1994.

[36] N. H. Frijda and others, *The emotions*. Cambridge University Press, 1986.

[37] R. K. Gupta and Y. Yang, "CrystalFeel at SemEval-2018 Task 1: Understanding and Detecting Emotion Intensity using Affective Lexicons," in *In Proceedings of The 12th International Workshop on Semantic Evaluation*, 2018, pp. 256–263.

[38] M. Pantic and L. J. Rothkrantz, "Toward an Affect-sensitive Multimodal Human-computer Interaction," *Proc. IEEE*, vol. 91, no. 9, pp. 1370–1390, 2003.

[39] A. Metallinou, M. Wollmer, A. Katsamanis, F. Eyben, B. Schuller, and S. Narayanan, "Context-sensitive Learning for Enhanced Audiovisual Emotion Classification," *IEEE Trans. Affect. Comput.*, vol. 3, no. 2, pp. 184–198, 2012.

[40] J. C. Lin, C. H. Wu, and W. L. Wei, "Error Weighted Semi-coupled Hidden Markov Model for Audio-visual Emotion

Recognition," *IEEE Trans. Multimed.*, vol. 14, no. 1, pp. 142–156, 2012.

[41] B. Schuller, "Recognizing Affect from Linguistic Information in 3D Continuous Space," *IEEE Trans. Affect. Comput.*, vol. 2, no. 4, pp. 192–205, 2011.

[42] C. H. Wu and W. B. Liang, "Emotion Recognition of Affective Speech Based on Multiple Classifiers Using Acoustic-prosodic Information and Semantic Labels," *IEEE Trans. Affect. Comput.*, vol. 2, no. 1, pp. 10–21, 2011.

[43] O. AlZoubi, S. K. D'Mello, and R. A. Calvo, "Detecting naturalistic expressions of nonbasic affect using physiological signals," *IEEE Trans. Affect. Comput.*, vol. 3, no. 3, pp. 298–310, 2012.

[44] L. Kessous, G. Castellano, and G. Caridakis, "Multimodal emotion recognition in speech-based interaction using facial expression, body gesture and acoustic analysis," *J. Multimodal User Interfaces*, vol. 3, no. 1–2, pp. 33–48, 2010.

[45] M. Wollmer, M. Kaiser, F. Eyben, B. Schuller, and G. Rigoll, "LSTM-Modeling of continuous emotions in an audiovisual affect recognition framework," *Image Vis. Comput.*, vol. 31, no. 2, pp. 153–163, 2013.

[46] M. Glodek *et al.*, "Multiple Classifier Systems for the Classification of Audio-visual Emotional States," *Affect. Comput. Intell. Interact. Springer*, pp. 359–368, 2011.

[47] B. de Gelder, H. K. M. Meeren, R. Righart, J. den Stock, W. A. C. de Riet, and M. Tamietto, "Beyond the face: exploring rapid influences of context on face processing," *Prog. Brain Res.*, vol. 155, pp. 37–48, 2006.

[48] F. Lingenfelser, J. Wagner, J. Deng, R. Brueckner, B. Schuller, and E. André, "Asynchronous and event-based fusion systems for affect recognition on naturalistic data in comparison to conventional approaches," *IEEE Trans. Affect. Comput.*, vol. 9, no. 4, pp. 410–423, 2016.

[49] J. C. Kim and M. A. Clements, "Multimodal Affect Classification at Various Temporal Lengths," *IEEE Trans. Affect. Comput.*, vol. 6, no. 4, pp. 371–384, 2015.

[50] H. Aviezer *et al.*, "Angry, Disgusted, or Afraid? Studies on the Malleability of Emotion Perception," *Psychol. Sci.*, vol. 19, no. 7, pp. 724–732, 2008.

[51] J. Brebner, "Gender and emotions," *Pers. Individ. Dif.*, vol. 34, no. 3, pp. 387–394, 2003.

[52] E. Hampson, S. M. van Anders, and L. I. Mullin, "A female advantage in the recognition of emotional facial expressions: Test of an evolutionary hypothesis," *Evol. Hum. Behav.*, vol. 27, no. 6, pp. 401–416, 2006.

[53] E. B. McClure, "A Meta-analytic Review of Sex Differences in Facial Expression Processing and their Development in Infants, Children, and Adolescents," *Psychol. Bull.*, vol. 126, pp. 424–453, 2000.

[54] L. Al-Shawaf, D. M. G. Lewis, and D. M. Buss, "Sex differences in disgust: Why are women more easily disgusted than men?," *Emot. Rev.*, vol. 10, no. 2, pp. 149–160, 2018.

[55] W. A. Babchuk, R. B. Hames, and R. A. Thompson, "Sex differences in the Recognition of Infant Facial Expressions of Emotion: The Primary Caretaker Hypothesis," *Ethol. Sociobiol.*, vol. 6, pp. 89–101, 1985.

[56] D. Keltner and J. Haidt, "Social Functions of Emotions at Four Levels of Analysis," *Cogn. Emot.*, vol. 13, no. 5, pp. 505–521, 1999.

[57] P. Barros, N. Churamani, E. Lakomkin, H. Siqueira, A. Sutherland, and S. Wermter, "The Omg-emotion Behavior Dataset.," in *2018 International Joint Conference on Neural Networks (IJCNN)*, 2018, pp. 1–7.

[58] V. Vonikakis, R. Subramanian, J. Arnfred, and S. Winkler, "Probabilistic Approach to People-centric Photo selection and Sequencing," *IEEE Trans. Multimed.*, vol. 19, no. 11, pp. 2609–2624, 2017.

[59] S. Peng, L. Zhang, S. Winkler, and M. Winslett, "Give me One Portrait Image, I will tell you your Emotion and Personality.," in *Proceedings of the 26th ACM International Conference on Multimedia*, 2018, pp. 1226–1227.

[60] V. Vonikakis and S. Winkler, "Emotion-based Sequence of Family Photos," in *Proceedings of the 20th ACM international conference on Multimedia*, 2012, pp. 1371–1372.

[61] V. Vonikakis, Y. Yazici, V. D. Nguyen, and S. Winkler, "Group Happiness Assessment using Geometric Features and Dataset Balancing," in *Proceedings of the 18th ACM International Conference on Multimodal Interaction*, 2016, pp. 479–486.

[62] H. W. Ng, V. D. Nguyen, V. Vonikakis, and S. Winkler, "Deep Learning for Emotion Recognition on Small Datasets using Transfer Learning," in *Proceedings of the ACM International Conference on Multimodal Interaction*, 2015, pp. 443–449.

[63] A. Mollahosseini, B. Hasani, and M. H. Mahoor, "Affectnet: A database for facial expression, valence, and arousal computing in the wild," *IEEE Trans. Affect. Comput.*, vol. 10, no. 1, pp. 18–31, 2017.

[64] D.-Y. Huang *et al.*, "Visual Speech Emotion Conversion using Deep Learning for 3D Talking Head," in *Proceedings of the Joint Workshop of the 4th Workshop on Affective Social Multimedia Computing and First Multi-Modal Affective Computing of Large-Scale Multimedia Data*, 2018, pp. 7–13.

[65] D. Y. Huang *et al.*, "Multimodal Prediction Of Affective Dimensions Fusing Multiple Regression Techniques," in *INTERSPEECH 2017*, 2017, pp. 162–165.

[66] X. Ouyang *et al.*, "Audio-visual Emotion Recognition Using Deep Transfer Learning and Multiple Temporal Models," in *19th ACM International Conference on Multimodal Interaction*, 2017, pp. 577–582.

[67] W. Ding *et al.*, "Audio and Face Video Emotion Recognition in the Wild using Deep Neural Networks and Small Datasets," in *Proceedings of the 18th ACM International Conference on Multimodal Interaction*, 2016, pp. 506–513.

[68] W. Y. Quck, D. Y. Huang, W. Lin, H. Li, and M. Dong, "Mobile Acoustic Emotion Recognition," in *IEEE Region 10 Conference (TENCON)*, 2016, pp. 170–174.

[69] K. Poon-Feng, D.-Y. Huang, M. Dong, and H. Li, "Acoustic emotion recognition based on fusion of multiple feature-dependent deep Boltzmann machines," in *The 9th International Symposium on Chinese Spoken Language Processing*, 2014, pp. 584–588.

[70] B. Schuller, M. Valstar, F. Eyben, G. McKeown, R. Cowie, and M. Pantic, "Avec 2011--the first international audio/visual emotion challenge," in *International Conference*

on Affective Computing and Intelligent Interaction, 2011, pp. 415–424.

[71] J. R. J. Fontaine, K. R. Scherer, E. B. Roesch, and P. C. Ellsworth, "The world of emotions is not two-dimensional," Psychol. Sci., vol. 18, no. 12, pp. 1050–1057, 2007.

[72] J. Dennis and T. H. Dat, "Single and multi-channel approaches for distant speech recognition under noisy reverberant conditions: I2R'S system description for the ASpIRE challenge," in 2015 IEEE Workshop on Automatic Speech Recognition and Understanding (ASRU), 2015, pp. 518–524.

[73] V. H. Do, N. F. Chen, B. P. Lim, and M. Hasegawa-Johnson, "Multi-task learning using mismatched transcription for under-resourced speech recognition," in Proceedings of the Annual Conference of the International Speech Communication Association, INTERSPEECH, 2017, pp. 734–738.

[74] R. Tong, N. F. Chen, and B. Ma, "Multi-Task Learning for Mispronunciation Detection on Singapore Children's Mandarin Speech," in INTERSPEECH, 2017, pp. 2193–2197.

[75] R. Duan and N. F. Chen, "Unsupervised Feature Adaptation Using Adversarial Multi-Task Training for Automatic Evaluation of Children's Speech," in INTERSPEECH, 2020, pp. 3037–3041.

[76] R. Duan and N. F. Chen, "Senone-aware Adversarial Multi-task Training for Unsupervised Child to Adult Speech Adaptation," in International Conference on Acoustics, Speech, & Signal Processing (ICASSP), 2021.

[77] N. F. Chen et al., "Strategies for Vietnamese keyword search," in IEEE International Conference on Acoustics, Speech and Signal Processing (ICASSP), 2014, pp. 4121–4125.

[78] M. Nguyen, G. H. Ngo, and N. Chen, "Multimodal neural pronunciation modeling for spoken languages with logographic origin," in Proceedings of the 2018 Conference on Empirical Methods in Natural Language Processing, 2018, pp. 2916–2922.

[79] M. Nguyen, G. H. Ngo, and N. F. Chen, "Hierarchical Character Embeddings: Learning Phonological and Semantic Representations in Languages of Logographic Origin Using Recursive Neural Networks," IEEE/ACM Trans. Audio, Speech, Lang. Process., vol. 28, pp. 461–473, 2019.

[80] C. Ni, C.-C. Leung, L. Wang, N. F. Chen, and B. Ma, "Unsupervised data selection and word-morph mixed language model for tamil low-resource keyword search," in 2015 IEEE International Conference on Acoustics, Speech and Signal Processing (ICASSP), 2015, pp. 4714–4718.

[81] S. M. Mohammad and F. Bravo-Marquez, "Emotion Intensities in Tweets," in Proceedings of the Sixth Joint Conference on Lexical and Computational Semantics, 2017, pp. 65–77.

[82] R. K. Gupta, P. Bhattacharya, and Y. Yang, "What Constitutes Happiness? Predicting and Characterizing the Ingredients of Happiness Using Emotion Intensity Analysis," in The AAAI-19 Workshop on Affective Content Analysis (AFFCON2019), 2019.

[83] R. K. Gupta and Y. Yang, "Predicting and Understanding News Social Popularity with Emotional Salience Features," in Proceedings of the 27th ACM International Conference on Multimedia, 2019, pp. 139–147.

[84] M. Soleymani, M. Pantic, and T. Pun, "Multimodal Emotion Recognition in Response to Videos," IEEE Trans. Affect. Comput., vol. 3, no. 2, pp. 211–223, 2012.

[85] N. V Chawla, K. W. Bowyer, L. O. Hall, and W. P. Kegelmeyer, "SMOTE: synthetic minority over-sampling technique," J. Artif. Intell. Res., vol. 16, pp. 321–357, 2002.

[86] P. Barros, D. Jirak, C. Weber, and S. Wermter, "Multimodal Emotional State Recognition Using Sequence-dependent Deep Hierarchical Features," Neural Networks, vol. 72, pp. 140–151, 2015.

[87] C. Busso et al., "Analysis of Emotion Recognition using Facial Expressions, Speech and Multimodal Information," in Proceedings of the 6th International Conference on Multimodal Interfaces, 2004, pp. 205–211.

[88] C. M. Lee et al., "Emotion Recognition Based on Phoneme Classes," in Eighth International Conference on Spoken Language Processing, 2004, pp. 889–892.

[89] H. Cao, D. G. Cooper, M. K. Keutmann, R. C. Gur, A. Nenkova, and R. Verma, "CREMA-D: Crowd-sourced emotional multimodal actors dataset," IEEE Trans. Affect. Comput., vol. 5, no. 4, pp. 377–390, 2014.

[90] L. R. Brody and J. A. Hall, "Gender and emotion in context," Handb. Emot., vol. 3, pp. 395–408, 2008.

[91] J. P. Bajorek, "Voice recognition still has significant race and gender biases," Harvard Business Review, 2019. [Online]. Available: https://hbr.org/2019/05/voice-recognition-still-has-significant-race-and-gender-biases.

[92] M. R. Costa-jussà, "An analysis of gender bias studies in natural language processing," Nat. Mach. Intell., vol. 1, no. 11, pp. 495–496, 2019.

**Prasanta Bhattacharya** is a Scientist at the Social and Cognitive Computing Department of the Institute of High Performance Computing, A*STAR, Singapore. His research interests cover social network analyses and behavioral analytics. He graduated with a PhD in Information Systems from the National University of Singapore. He also serves as an Adjunct Assistant Professor with the Department of Analytics and Operations at NUS Business School.

**Raj Kumar Gupta** is a Senior Scientist at the Social and Cognitive Computing Department of the Institute of High Performance Computing, A*STAR, Singapore. His research interests include natural language processing, sentiment and emotion analysis, and multimodal analysis. He received his PhD in Computer Science from the Nanyang Technological University, Singapore. He has more than five years' industry experience on software development, visual analytics and embedded systems.

**Yinping Yang** is a Senior Scientist and Principal Investigator of the A*STAR Digital Emotions programme at Institute of High Performance Computing, A*STAR, Singapore. Her research interests cover intelligent negotiation systems, emotion recognition and sentiment analysis, and strategic foresight. She obtained her Ph.D. in Information Systems from the National University of Singapore. Besides research, she also manages industry projects and advises tech start-ups in emotion analytics and e-negotiation technologies.

# Exploring the contextual factors affecting multimodal emotion recognition in videos

**Supplemental Material**

**Appendix A**         **Multimodal Emotion Features and Correlation Analyses with OMG Emotion Ground Truth Labels**

Appendix A provides a detailed summary of the 43 visual, audio and text based emotion features, and reports results from a correlation analysis. This supplemental material complements the feature extraction description reported in Section 4 of the main paper.

Prior to our classification and regression experiments, we performed a bivariate correlation analysis to gain a preliminary understanding of how each of the extracted visual, audio and text features is associated with the ground truth emotion labels in the OMG videos dataset. As the emotion classes and most of the features are binary or dichotomous variables, Kendall's correlation coefficient $\tau$ was used for all correlations, except for the text-based features which are continuous variables, for which Pearson's $r$ was used.

Table A1 below presents the results from the bivariate correlation analysis. The results suggest a clear pattern: at the individual feature level, almost all the extracted visual, audio, and text based emotion features were significantly associated with at least one of the seven emotion classes.

Among the visual features, *vValence* ($\tau$ = .343\*\*) and *vIntensity* ($\tau$ = .252\*\*) were found to have the strongest correlations with happiness. For audio features, *aArousal* ($\tau$ = .141\*\*) and *aIntensity* ($\tau$ = .143\*\*) were significantly correlated with anger. For text based emotion features, *tValence*, i.e., the intensity of the overall valence detected in the speech transcript, was found to be highly associated with happiness ($r$ = .163\*\*). Meanwhile, *tSadness*, i.e., the intensity of sadness detected in the speech transcript, was highly associated with sadness ($r$ = .164\*\*).

The correlation analysis also indicated that gender and duration were indeed associated with some emotions, but not others in this dataset. For example, female speakers were negatively associated with neutral ($\tau$ = -.214\*\*) expressions, but positively associated with sadness ($\tau$ = .256\*\*) and anger ($\tau$ = .046\*) expressions. Longer videos were positively associated with happiness ($\tau$ = .089\*\*) and surprise ($\tau$ = .044\*) but were negatively associated with sadness ($\tau$ = -.072\*\*).

The correlation results revealed interesting preliminary patterns, but most of the above-illustrated correlations have relatively small magnitudes. In order to understand the predictive performance of these features, we used machine learning based classifiers, as reported in the main paper, to analyze the combined effects of these various individual visual, audio and text based features.

## Table A1. Results of Bivariate Correlation Analysis

| Feature Name | Scale Range and Description | Correlation Coefficients (Kendall's τ) | | | | | | |
|---|---|---|---|---|---|---|---|---|
| | | Neutral | Happiness | Sadness | Anger | Disgust | Fear | Surprise |
| **Visual-based Emotion Features+** | | | | | | | | |
| *vValence* | A continuous variable indicating the degree to which the facial expressions express [-1: most unpleasant feelings; 1: most pleasant feelings] | **-.152**** | **.343**** | **-.119**** | **-.089**** | -.027 | **-.057**** | .011 |
| *vArousal* | A continuous variable indicating the degree to which the facial expressions have [-1: no heightened physiological activity; 1: highest degree of heightened physiological activity] | **.051*** | .027 | **-.195**** | **.082**** | .022 | -.029 | **.049*** |
| *vIntensity* | A continuous variable indicating the degree to which the facial expressions express [0: least intense feelings; 1: most intense feelings] | **-.080**** | **.252**** | **-.184**** | -.019 | -.008 | **-.050*** | .029 |
| *vAlarmed* | A group of binary variables indicating whether or not the facial expressions express this emotional state [0: no; 1: yes] | **.061**** | -.020 | **-.043*** | .014 | -.024 | -.007 | -.028 |
| *vAroused* | | .025 | **-.055*** | -.021 | **.047*** | .022 | -.010 | .021 |
| *vAstonished* | | **.052*** | .003 | **-.082**** | -.006 | .017 | -.011 | .025 |
| *vExcited* | | .014 | .008 | **-.067**** | .033 | -.010 | -.004 | **.048*** |
| *vFrustrated* | | .013 | **-.068**** | **.090**** | .021 | -.036 | -.026 | -.017 |
| *vGloomy* | | -.025 | -.037 | .023 | .004 | .026 | **.044*** | **.072**** |
| *vHappy* | | **-.049*** | **.139**** | -.034 | **-.064**** | -.001 | -.040 | -.009 |
| *vMiserable* | | -.021 | **-.073**** | **.141**** | -.023 | .009 | .005 | -.013 |
| *vNeutral* | | -.029 | **-.072**** | **.185**** | **-.052*** | -.010 | .019 | -.019 |
| *vPleased* | | **-.119**** | **.235**** | **-.058**** | **-.065**** | -.026 | -.015 | -.016 |
| *vTensed* | | -.005 | **-.045*** | **-.047*** | **.099**** | .026 | .021 | -.014 |
| **Audio-based Emotion Features** | | | | | | | | |
| *aValence* | A continuous variable indicating the degree to which the audio expresses [-1: most unpleasant feelings; 1: most pleasant feelings] | .021 | -.011 | .032 | **-.058**** | .011 | .008 | -.007 |
| *aArousal* | A continuous variable indicating the degree to which the audio has [-1: no heightened physiological activity; 1: highly heightened physiological activity] | **-.095**** | **-.043*** | .007 | **.141**** | **.073**** | -.033 | **.066**** |
| *aPower* | A continuous variable indicating the degree to which the audio expresses [-1: least sense of power; 1: most sense of power] | -.040 | **.079**** | **-.103**** | **.060**** | .030 | **-.050*** | .009 |

| Emotion Features | Emotion Feature description | Correlation Coefficients (Kendall's τ) | | | | | | |
| --- | --- | --- | --- | --- | --- | --- | --- | --- |
| | | Neutral | Happiness | Sadness | Anger | Disgust | Fear | Surprise |
| *aExpectancy* | A continuous variable indicating the degree to which the audio expresses [-1: least sense of expectancy; 1: most sense of expectancy] | -.032 | **.044*** | -.007 | .003 | -.018 | -.033 | **.051*** |
| *aIntensity* | A continuous variable indicating the degree to which the audio expresses [0: least intense feelings; 1: most intense feelings] | **-.065**** | **.063**** | **-.141**** | **.143**** | .034 | **-.063**** | **.074**** |
| *aHappiness* | A group of binary variables indicating whether or not the audio expresses this emotion [0: no; 1: yes] | **-.044*** | -.029 | **.066**** | .018 | .013 | .019 | -.002 |
| *aAnger* | | -.039 | .001 | .007 | **.045*** | .036 | -.037 | -.005 |
| *aFear* | | .033 | **-.074**** | .026 | .037 | -.029 | .032 | -.013 |
| *aSadness* | | .032 | **.077**** | **-.044*** | **-.091**** | -.012 | .000 | -.027 |
| *aNeutral* | | -.018 | -.001 | **-.049*** | **.062**** | .018 | -.040 | **.084**** |
| **Text-based Emotion Features** | | | | | | | | |
| *tValence* | A continuous variable indicating the degree to which the text expresses [0: highly intense unpleasant feelings; 1: highly intense pleasant feelings] | **.050*** | **.163**** | **-.130**** | **-.133**** | -.037 | -.013 | -.013 |
| *tJoy* | A group of continuous variables indicating the degree to which the text expresses [0: barely noticeable amount of joy/anger/fear/sadness; 1: extremely high intensity of joy/anger/fear/sadness] | .000 | **.161**** | -.098 | **-.102**** | -.025 | -.014 | -.001 |
| *tAnger* | | **-.050*** | -.058 | -.006 | **.129**** | .032 | .007 | .028 |
| *tFear* | | **-.055*** | **-.114**** | **.107**** | **.112**** | .003 | .021 | .015 |
| *tSadness* | | **-.082**** | **-.145**** | **.164**** | **.145**** | -.002 | .008 | .015 |
| **Contextual Factors** | | | | | | | | |
| **female** | Binary variable indicating whether or not the speaker in the video is female [1: yes; 0: no] | **-.214**** | -.004 | **.256**** | **.046*** | .004 | -.015 | .008 |
| **male** | Binary variable indicating whether or not the speaker in the video is male [1: yes; 0: no] | **.148**** | **-.138**** | **-.083**** | .035 | .026 | .038 | -.013 |
| **others** | Binary variable indicating whether or not the speaker in the video is specific gender-identifiable [1: yes; 0: no] | **.113**** | **.140**** | **-.192**** | **-.117**** | -.033 | **-.049*** | .013 |
| **duration** | The duration of emotion expression episode measured in seconds | -.040 | **.089**** | **-.072**** | .019 | -.004 | -.040 | **.044*** |

** Correlation is significant at the 0.01 level. * Correlation is significant at the 0.05 level.
+ For visual-based features, only those individual features with more than one significant correlation were listed here, due to space limitations.

# Appendix B    Additional Classifier Experiment Results

Appendix B provides results from additional analyses which we performed to test and confirm the robustness of the findings reported in Section 5 of the main paper.

Table B1 presents AUC scores for our focal models using a random forest classifier with the i) number of trees and ii) number of variables randomly sampled at each split (i.e. the ntree and mtry parameters) tuned using a grid-search approach that minimizes out-of-bag (OOB) error.

**Table B1. Performance analysis using a random forest classifier with class weights (cross-validated AUC scores)**

| Modality | AUC | | | |
|---|---|---|---|---|
| | Neutral | Happiness | Sad | Anger |
| Visual | **.576** | **.660** | .693 | .555 |
| Audio | .553 | .570 | .651 | **.601** |
| Text | .558 | .600 | .655 | .581 |
| V + A | **.613** | .705 | .755 | **.661** |
| V + T | .608 | **.712** | **.756** | .624 |
| A + T | .578 | .635 | .726 | .625 |
| V + A + T | **.620** | **.721** | **.795** | **.663** |
| MM1 Score | 7.6 | 9.2 | 14.7 | 10.3 |

In Table B2, we present results from a SMOTE analysis to check for robustness to class imbalance. In the previous analysis, class imbalance was addressed by assigning class weights to the classifiers, proportional to the class distributions. Next, we present results using Synthetic Minority Oversampling Technique (SMOTE)[1], which oversamples from the minority class not by repeat sampling but through synthesis of additional observations in the minority class. The SMOTE synthesis technique works by first randomly selecting an instance from the minority class, and choosing its $k$ nearest neighbors from the same class (based on a tunable parameter $k$). Next, it selects a random instance from this new list of neighbors, and synthesizes a new instance on the feature space enclosed between these two selected instances. Studies have shown that under-sampling of the majority class together with SMOTE-based synthesis can offer superior classifier performance on the AUC metric.

It is important to note that we integrate the SMOTE oversampling approach with our cross-validation (CV) step to perform the classification process. Blagus and Lusa (2015)[2] points out SMOTE-augmenting the entire dataset before performing CV often leads to inflated CV estimates, and that this should be avoided. Hence, in our study, we perform SMOTE only on the specific folds used for training, while leaving the testing folds untouched. The results from our SMOTE analysis are presented in Table B2 below. For this analysis, we synthesized 2 new observations for each minority class observation (i.e., *over-sampling parameter* = 2), and with $k$=2 (i.e. 2 nearest neighbors). The number of majority class observations that were randomly selected to be retained in the final sample was equal to the number of newly synthesized minority class observations (i.e., *under-sampling parameter* = 1)[3].

---

[1] Chawla, N. V., Bowyer, K. W., Hall, L. O., & Kegelmeyer, W. P. (2002). SMOTE: synthetic minority over-sampling technique. *Journal of artificial intelligence research*, *16*, 321-357.

[2] Lusa, L. (2015). Joint use of over-and under-sampling techniques and cross-validation for the development and assessment of prediction models. *BMC bioinformatics*, *16*(1), 1-10.

**Table B2. Performance analysis using a random forest classifier with SMOTE technique**

| Modality | AUC (Cross-Validated) | | | |
| --- | --- | --- | --- | --- |
| | *Neutral* | *Happiness* | *Sad* | *Anger* |
| *Visual* | **.560** | **.617** | **.631** | .534 |
| *Audio* | .536 | .566 | .606 | **.566** |
| *Text* | .532 | .554 | .614 | .545 |
| *V + A* | **.565** | **.629** | .662 | **.594** |
| *V + T* | .557 | **.629** | **.676** | .576 |
| *A + T* | .555 | .594 | .664 | .588 |
| *V + A + T* | **.589** | **.636** | **.692** | **.610** |
| *MM1 Score (%)* | 5.2 | 3.1 | 9.7 | 7.8 |

In Table B3 below, we provide the distribution of the neutral, happiness, sadness, and anger emotion categories across the various gender and duration subgroups.

**Table B3. Sample description for gender and duration subgroups**

| Subgroups | Neutral | | Happiness | | Sadness | | Anger | |
| --- | --- | --- | --- | --- | --- | --- | --- | --- |
| | 0 | 1 | 0 | 1 | 0 | 1 | 0 | 1 |
| *Male* | 412 | 326 | 571 | 167 | 670 | 68 | 633 | 105 |
| *Female* | 759 | 245 | 705 | 299 | 761 | 243 | 868 | 136 |
| *Short Duration* | 325 | 220 | 419 | 126 | 451 | 94 | 480 | 65 |
| *Long Duration* | 354 | 189 | 360 | 183 | 474 | 69 | 480 | 63 |

Tables B4-B7 present performance measures for gender (male vs. female) and duration (short duration vs. long duration) subgroups using a random forest classifier.

**Table B4. Performance analysis for male subgroup using a random forest classifier**

| Modality | AUC (Cross-Validated) | | | |
| --- | --- | --- | --- | --- |
| | *Neutral* | *Happiness* | *Sad* | *Anger* |
| *Visual* | **.552** | **.607** | **.770** | **.591** |
| *Audio* | .548 | .556 | .702 | .565 |
| *Text* | .511 | .538 | .546 | .543 |
| *V + A* | **.578** | **.645** | **.848** | **.637** |
| *V + T* | .543 | .639 | .826 | .608 |
| *A + T* | .563 | .571 | .731 | .576 |
| *V + A + T* | **.599** | **.637** | **.850** | **.642** |
| *MM1 Score* | 8.5 | 6.3 | 10.4 | 8.6 |
| *MM1 Score (Overall)* | 7.6 | 9.2 | 14.7 | 10.3 |

---

[3] We replicated the SMOTE analysis with different over-sampling parameters and nearest neighbors (i.e. *over-sampling* = [3,4], and *k*=[3, 4, 5]). The results from these analyses were consistent with the ones reported in Table B3.

**Table B5. Performance analysis for female subgroup using a random forest classifier**

| Modality | AUC (Cross-Validated) | | | |
|---|---|---|---|---|
| | Neutral | Happiness | Sad | Anger |
| Visual | **.591** | **.780** | **.741** | **.646** |
| Audio | .565 | .588 | .702 | .587 |
| Text | .521 | .609 | .618 | .585 |
| V + A | .606 | **.789** | **.794** | .658 |
| V + T | **.609** | **.789** | .769 | **.660** |
| A + T | .563 | .656 | .747 | .629 |
| V + A + T | **.613** | **.795** | **.809** | **.689** |
| MM1 Score | 3.7 | 1.9 | 9.2 | 6.7 |
| MM1 Score (Overall) | 7.6 | 9.2 | 14.7 | 10.3 |

**Table B6. Performance analysis for short duration subgroup using a random forest classifier**

| Modality | AUC (Cross-Validated) | | | |
|---|---|---|---|---|
| | Neutral | Happiness | Sad | Anger |
| Visual | .529 | **.726** | **.672** | .469 |
| Audio | **.570** | .548 | .638 | **.603** |
| Text | .538 | .574 | .601 | .513 |
| V + A | .570 | .747 | **.722** | .623 |
| V + T | **.580** | **.749** | .682 | .557 |
| A + T | .562 | .566 | .713 | **.644** |
| V + A + T | **.589** | **.765** | **.733** | **.676** |
| MM1 Score | 3.3 | 5.4 | 9.1 | 12.1 |
| MM1 Score (Overall) | 7.6 | 9.2 | 14.7 | 10.3 |

**Table B7. Performance analysis for long duration subgroup using a random forest classifier**

| Modality | AUC (Cross-Validated) | | | |
|---|---|---|---|---|
| | Neutral | Happiness | Sad | Anger |
| Visual | **.610** | **.630** | .774 | **.630** |
| Audio | .527 | .539 | .571 | .564 |
| Text | .535 | .600 | .681 | .532 |
| V + A | **.627** | .662 | .778 | .654 |
| V + T | .619 | **.686** | **.821** | .634 |
| A + T | .511 | .627 | .693 | **.672** |
| V + A + T | **.642** | **.690** | **.821** | **.680** |
| MM1 Score | 5.2 | 9.5 | 6.1 | 7.9 |
| MM1 Score (Overall) | 7.6 | 9.2 | 14.7 | 10.3 |